\documentclass[runningheads]{svmult}

\usepackage{makeidx}   
\usepackage{graphicx}  
\usepackage{subeqnar}  
\usepackage{multicol}  
\usepackage{physprbb}  
\makeindex             

\newcommand{\sax}{{\em Beppo}SAX}
\newcommand{\grb}{GRB000615}

\def\deg{\ifmmode^{\circ}\else$^{\circ}$\fi} 
\def\farcm{\hbox{$\,.\!\!^{\prime}$}}

\def\fs{\hbox{$\,.\!\!^{\rm s}$}}
\def\lapp{\ifmmode\stackrel{<}{_{\sim}}\else$\stackrel{>}{_{\sim}}$\fi}
\def\gapp{\ifmmode\stackrel{>}{_{\sim}}\else$\stackrel{>}{_{\sim}}$\fi}

\newcommand{\fu}{erg~cm$^{-2}$ s$^{-1}$}

\begin{document}
\title*{\grb\ in X-rays}
\toctitle{\grb\ in X-rays}
%
%
\titlerunning{\grb\ in X-rays}
%
\author{Luciano Nicastro\inst{1}
\and G. Cusumano\inst{1}
\and A.~Antonelli\inst{2}
\and L. Amati\inst{3}
\and F. Frontera\inst{3}
\and E. Palazzi\inst{3}
\and E. Pian\inst{4}
\and E. Costa\inst{5}
\and M. Feroci\inst{5}
\and L. Piro\inst{5}
\and J. in 't Zand\inst{6}
\and J. Heise\inst{6}
}
\authorrunning{Luciano Nicastro et al.}
%
%
\institute{IFCAI -- CNR, Via U. La Malfa 153, 90146 Palermo (I)
\and Osservatorio Astr. di Roma, Via Frascati 33, 00040
 Monteporzio Catone, Roma (I)
\and Osservatorio Astr. di Trieste, Via G.B. Tiepolo, 11, 34131 Trieste (I)
\and ITeSRE -- CNR, Via P. Gobetti 101, 40129 Bologna (I)
\and IAS -- CNR, Via Fosso del Cavaliere, 00131 Roma (I)
\and SRON, Sorbonnelaan 2, 3584 CA Utrecht (NL)
}

\maketitle              

\begin{abstract}
\index{abstract}
\grb\ was detected simultaneously by the \sax\ GRBM and WFC 1 with a
localization uncertainty of $2'$ (error circle radius).
X-ray emission was detected only in the 0.1--4 keV range during a NFI
observation started $\simeq 10$ hours after the trigger time.  The
positional and temporal analysis shows the presence of two sources, one of
which may be related to the GRB.
\end{abstract}

\section{Introduction}
\grb\ was detected on 15 June
2000, 06:17:44.91 UT \cite{piro1} at coordinates (J2000)
R.A. = $15^{\rm h}\, 32^{\rm m}\, 36\fs9$,
Dec  = $+73\deg\, 49'\, 07''$ ($2'$ error radius),
which are revised with respect to those distributed in GCN 705~\cite{gandolfi1}.
\sax\ NFI observation started the same day at 16:20:14 UT, i.e.
$\simeq 10$ hours
after the trigger time, and lasted 1.44 days. The net exposure time was
44609 s for MECS and 31280 s for LECS.
Two distinct uncatalogued X-ray sources are detected, one for each instrument.
The source detected in the MECS has constant flux during the observation
and its position is only marginally compatible with the WFC position
of the GRB. The LECS source (0.1--4 keV), though detected at low statistical
significance, is visible only in the first half of the observation;
its position is fully compatible with the WFC one.

Optical/IR/radio searches did not detect any new source
(see GCN circulars 706, 708, 709, 713, 719, 721, 727).

\section{The burst}
The GRBM (40--700 keV) light curve (lasting
${\bf \simeq 13}$ s) is shown in the top panel of Fig. 1.
Spectral analysis shows a softening during the burst with a power-law
spectral index (${\bf N_E\propto E^{-\alpha}}$)
from $1.0\pm 0.4$ (rise) to $2.4\pm 0.7$ (last part of the decay)
with an average of $\alpha=1.71 \pm 0.46$.
The average flux in the 40--700 keV band is
$f_\gamma = (8.2 \pm 0.7)\times 10^{-8}$ \fu,
the peak flux is $(1.0 \pm 0.2)\times 10^{-7}$ \fu\ and
the total fluence $F_\gamma = (9.8 \pm 0.9)\times10^{-7}$
erg cm$^{-2}$.
The hardness ratio $f_{100-300/50-100} = 2.0 \pm 0.3$.

\medskip
In the 2--28 keV WFC band, the burst behaviour is quite different.
At the time of the GRB onset, the X-ray flux increases only slightly, while
a major rise appears some 40 s later and lasts
$\simeq 60$ s.  Splitting the WFC band in the ranges
2--10 and 10--28 keV we note that the prompt emission comes mainly from
the hard band, while the delayed emission from the soft band.
This effect, which is under study, can be attributed to scattering of
gamma-rays to lower energies rather than to ``clean'' source photons.
The 2--28 keV (delayed) flux is $(2.85\pm 0.45)\times 10^{-8}$
\fu.
In 2--10 keV it is $f_x=(1.74 \pm 0.10)\times 10^{-8}$ \fu.

\section{\sax\ NFIs observation}

The follow-up \sax\ NFIs observation showed, in the MECS energy range
1.6--4 keV, a soft and stable X-ray source ($\sim 5\sigma$ level).
An optimized source counts extraction radius of $3'$ was used.
Figure 2 shows the smoothed image of the entire observation.
The source coordinates are: R.A. = $15^{\rm h}\, 32^{\rm m}\, 23\fs3$,
Dec = $+73\deg\, 47'\, 30''$ (1SAX J1532.4+7349).
To be \emph{conservative}, we can set an error circle of $2'$ (see Fig. 2).
This position (note it is $\simeq 1\farcm5$ away from that reported
in GCN 707~\cite{gandolfi2}) could then be compatible (offset $=1\farcm8$)
with the WFC
GRB position. However its flux stability suggests it is not related to the GRB.
The average flux is $f_x=(6.7 \pm 1.5)\times 10^{-14}$ \fu\ in
1.6--4 keV. No significant emission is detected above 4 keV.

In the LECS (0.1--4 keV) a source is detected at $2.5 \sigma$ level, but
\emph{only in the first half} of the observation.
The smoothed images of the two halves of the observation are shown in
Fig. 3.
The position of this source is: R.A. = $15^{\rm h}\, 32^{\rm m}\, 35$,
 Dec = $+73\deg\, 48'\, 50''$ (1SAX J1532.6+7348),
fully compatible with the WFC (GRB) position.
The low significance of the detection cannot help in discriminating the
source characteristics.

Unfortunately the low statistics leaves open several interpretations.
In particular it is not clear if the MECS and LECS sources are really distinct
and which (if any) is related to \grb.
More sensitive observations of the field may help investigating
the two detected sources.
Nevertheless, one can speculate that a soft prompt (actually \emph{delayed}) X-ray
event could lead to a soft afterglow.
In brief, our refined data analysis shows that:
\begin{itemize}
\item a stable (probably unrelated to the GRB) source is detected in the MECS.
 Its spectrum can be fit by a power-law of
 index $\alpha=3.2\pm 0.5$ ($\chi^2_n=1.1$).
\item after $\sim 20$ hours from the gamma-ray event the X-ray afterglow is not
 detected to the limit of the MECS sensitivity
 suggesting a power-law temporal decay index $\lapp -1.7$ (respect to the WFC
 mean flux and $T_{start}=T_{burst}$).
\item a source is detected in the LECS band 0.1--4 keV; some hint of decay is
 present.
\end{itemize}

\begin{figure}[h]
\label{lecsimg}
\begin{center}
\begin{tabular}{cc}
\includegraphics[width=5.80cm]{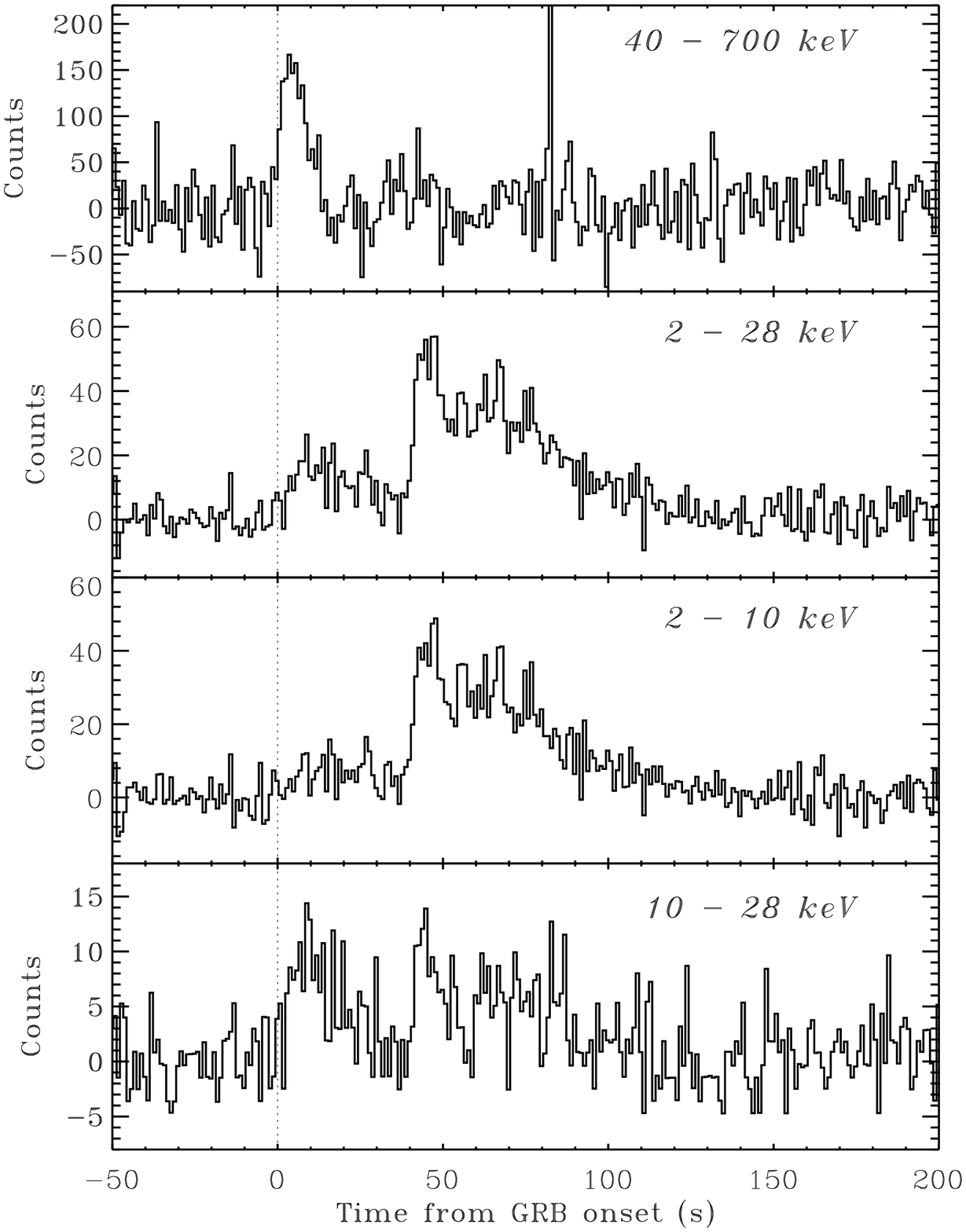} &
\includegraphics[width=5.80cm]{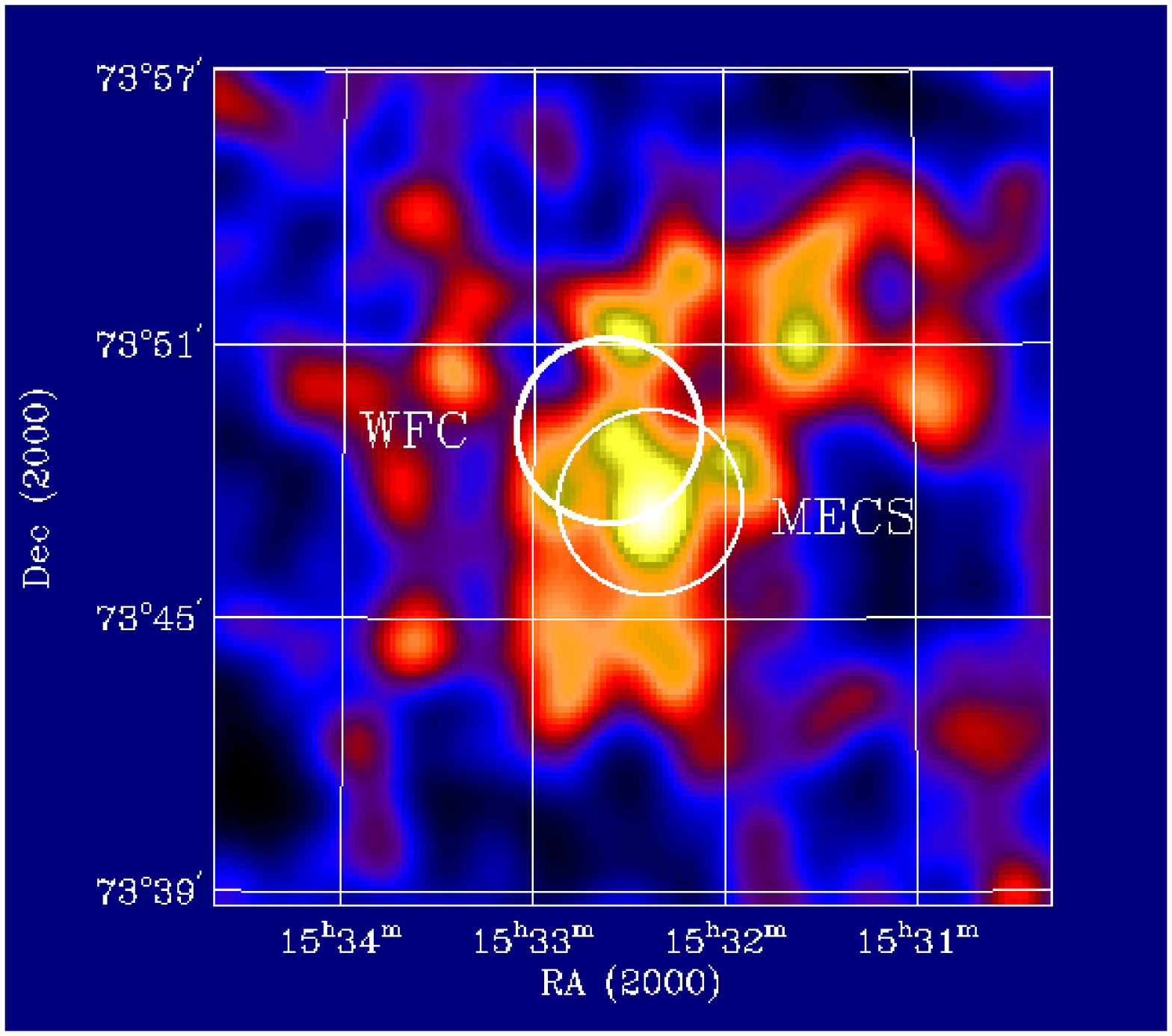} \\
\begin{minipage}{5.7cm}
{\bf Fig. 1.}
  The 1 s GRBM (\emph{top}) and WFC light curves.
 A hard X-ray component is visible in the 10--28 keV band at the GRB onset.
 The soft emission starts after $\simeq 40$ s from the trigger
\end{minipage} &
\begin{minipage}{5.7cm}
{\bf Fig. 2.}
 MECS 1.6--4 keV smoothed image.
 The $2'$ error circle is shown together with the WFC one (also $2'$).
 The main source is stable during the observation
\end{minipage} \\
 & \\
\includegraphics[width=5.80cm]{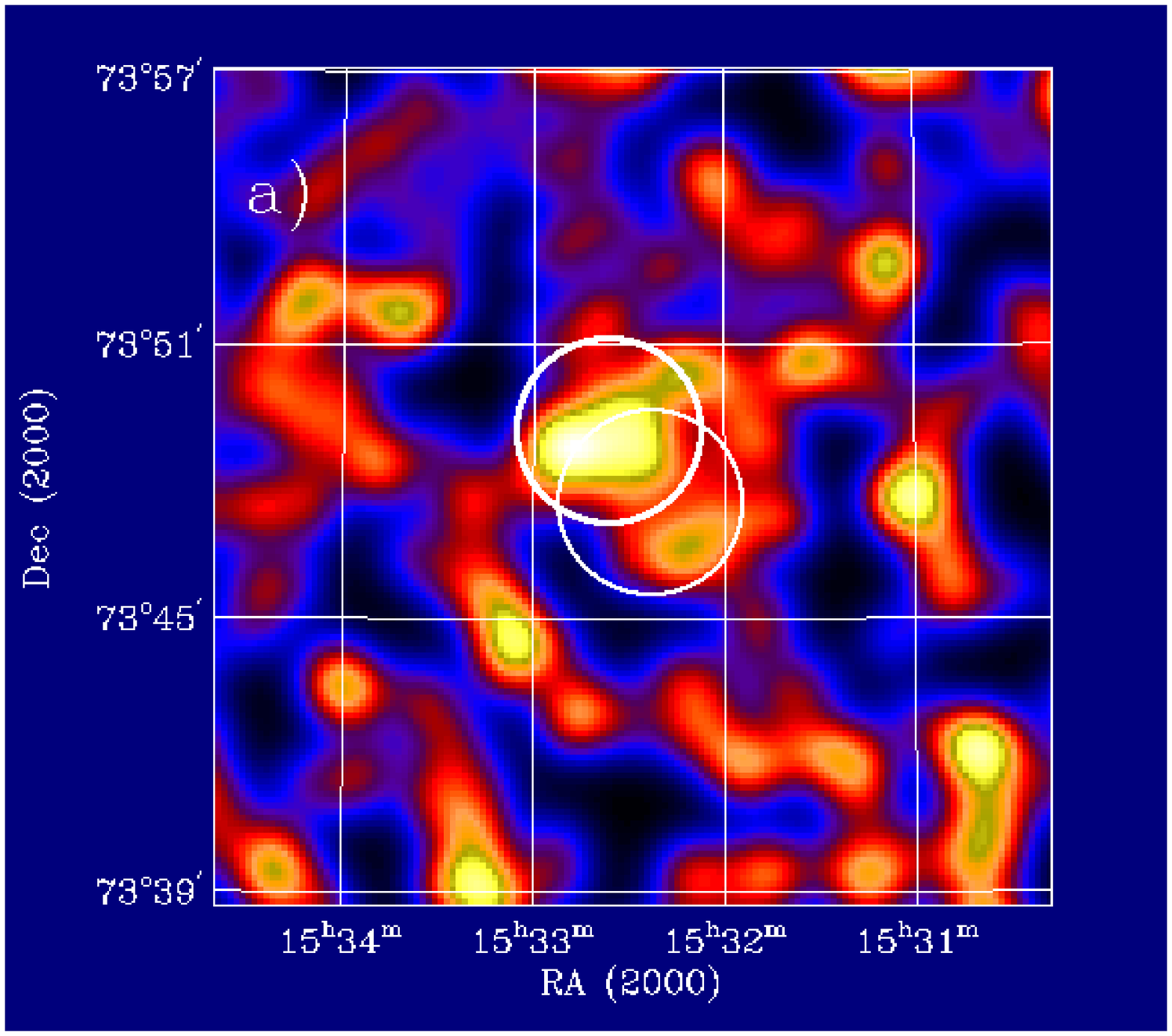} &
\includegraphics[width=5.80cm]{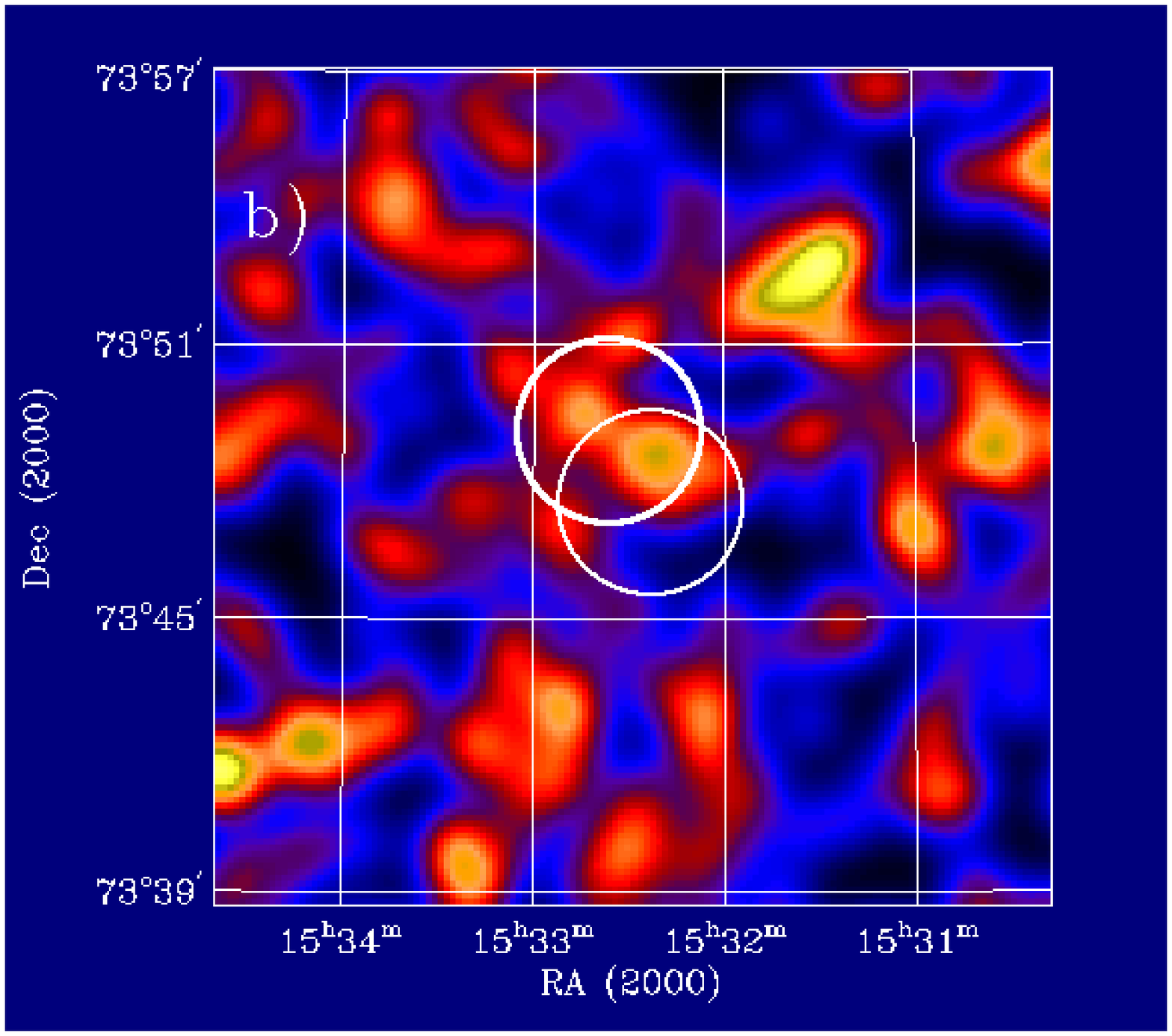} \\
  & \\
\multicolumn{2}{l}{
\begin{minipage}{11.7cm}
{\bf Fig. 3.}
 LECS (0.1--4 keV) images for the first ({\bf a}) and second ({\bf b})
 half of the observation. The WFC and MECS source error circles are shown
\end{minipage}
}
\end{tabular}
\end{center}
\end{figure}

%


\begin{thebibliography}{8.}
\addcontentsline{toc}{section}{References}

\bibitem{gandolfi1}
G. Gandolfi et al.: GCN circ. 705 (2000)

\bibitem{gandolfi2}
G. Gandolfi et al.: GCN circ. 707 (2000)

\bibitem{piro1}
L. Piro et al.: GCN circ. 703 (2000)

\end{thebibliography}
\end{document}